\documentclass[prd,twocolumn,showpacs,preprintnumbers,amsmath,amssymb,nofootinbib]{revtex4}

\usepackage{amsmath}
\usepackage{graphicx}
\usepackage{dcolumn}
\usepackage{xspace}
\usepackage{color}
\usepackage{url}
\usepackage[utf8]{inputenc}

\newcommand{\bea}{\begin{eqnarray}}
\newcommand{\eea}{\end{eqnarray}}
\newcommand{\nn}{\nonumber\\}

\newcommand{\eq}[1]{Eq.~\eqref{#1}}

\begin{document}
\preprint{PSI-PR-16-17, ZU-TH 44/16}
\title{$(g-2)_\mu$, Lepton Flavour Violation and $Z$ Decays with Leptoquarks:\\ Correlations and Future Prospects}

\author{Estefania Coluccio Leskow}
\email{coluccio@na.infn.it}
\affiliation{INFN-Sezione di Napoli, Via Cintia, 80126 Napoli, Italia}
\author{Andreas Crivellin}
\email{andreas.crivellin@cern.ch}
\affiliation{Paul Scherrer Institut, CH--5232 Villigen PSI, Switzerland}
\author{Giancarlo D'Ambrosio}
\email{gdambros@na.infn.it}
\affiliation{INFN-Sezione di Napoli, Via Cintia, 80126 Napoli, Italia}
\author{Dario M\"uller}
\email{dario.mueller@psi.ch}
\affiliation{Paul Scherrer Institut, CH--5232 Villigen PSI, Switzerland}
\affiliation{Physik-Institut, Universit\"at Z\"urich, Winterthurerstrasse 190, CH-8057 Z\"urich, Switzerland}

\begin{abstract}
The long-standing anomaly in the anomalous magnetic moment of the muon indicates the presence of chirality violating new physics contributions. A possible solution involves scalar leptoquarks with left- and right-handed couplings to the top quark. Two such representations of scalar leptoquarks exist for which the contribution to $(g-2)_\mu$ can possess an $m_t/m_\mu$ enhancement compared to the Standard Model. The leptoquarks also induce loop corrections to $Z$ couplings to muons which probe as well new physics contributions which possess sources of $SU(2)$ symmetry breaking and we find that this effect should be observable at future experiments as GigaZ or TLEP. Furthermore, once interactions of the leptoquark with tau leptons and electrons are present, additional correlated effects in anomalous magnetic moments, $Z\to\ell\ell^\prime$ and $\ell\to\ell^\prime\gamma$ arise, which can be used to test the model and to determine the flavour structure of the couplings. We find that the two representations of leptoquarks can be distinguished also from low energy experiments: one representation predicts constructive interference with the Standard Model in $Z$ couplings to leptons and effects in $B\to K^{(\star)}\nu\bar\nu$, while the other representation interferes destructively with the Standard Model in $Z$ couplings to leptons and gives a $C_9=C_{10}$-like contribution to $b\to s\ell^+\ell^-$ processes.
\end{abstract}

\maketitle

\section{Introduction}
\label{intro}

So far, the LHC did not directly observe any particles beyond the ones present in the Standard Model (SM) of particle physics. However, we have several hints for lepton flavour (universality) violating new physics (NP). Despite the anomalies in $B$ physics and the hints for a non-vanishing decay rate of $h\to\tau\mu$ (see Ref.~\cite{Crivellin:2016ekz} for an overview), the best known (and oldest) of these anomalies is the one in the anomalous  magnetic moment (AMM) of the muon. The world average of the measurement of $a_\mu \equiv (g-2)_\mu/2$ is completely dominated by the Brookhaven experiment E821~\cite{Bennett:2006fi} and is given by~\cite{Agashe:2014kda}
$a_\mu^\mathrm{exp} = (116\,592\,091\pm54\pm33) \times 10^{-11}$
where the first error is statistical and the second one is systematic. The current SM prediction is~\cite{Czarnecki1995,Czarnecki1996,Gnendiger2013,Davier2011,Hagiwara2011,Kurz2014,Jegerlehner2009,Aoyama2012,Colangelo2014,Nyffeler:2016gnb}
$	a_\mu^\mathrm{SM} = (116\,591\,811\pm62) \times 10^{-11}$
where almost the whole uncertainty is due to hadronic effects. This amounts to a discrepancy between the SM and the experimental value of 
\begin{equation}
	a_\mu^\mathrm{exp}-a_\mu^\mathrm{SM} = (278\pm 88)\times 10^{-11}\,,\label{muonAMMexp}
\end{equation}
i.e.~a $3.1\sigma$ deviation. Possible NP explanations besides supersymmetry (see for example Ref.~\cite{Stockinger:2006zn} for a review) include very light $Z^\prime$ bosons~\cite{Langacker:2008yv,Baek:2001kca,Ma:2001md,Gninenko:2001hx,Pospelov:2008zw,Heeck:2011wj,Harigaya:2013twa,Altmannshofer:2014pba,Altmannshofer:2016brv}, additional fermions~\cite{Freitas:2014pua}, new scalars~\cite{Iltan:2001nk,Broggio:2014mna,Wang:2014sda,Omura:2015nja,Crivellin:2015hha,Altmannshofer:2016oaq,Batell:2016ove}, or other vectors~\cite{Czarnecki:2001pv,Biggio:2014ela,Biggio:2016wyy}.

\begin{figure*}[tb]
\begin{center}
\begin{tabular}{cp{7mm}c}
\includegraphics[width=0.45\textwidth]{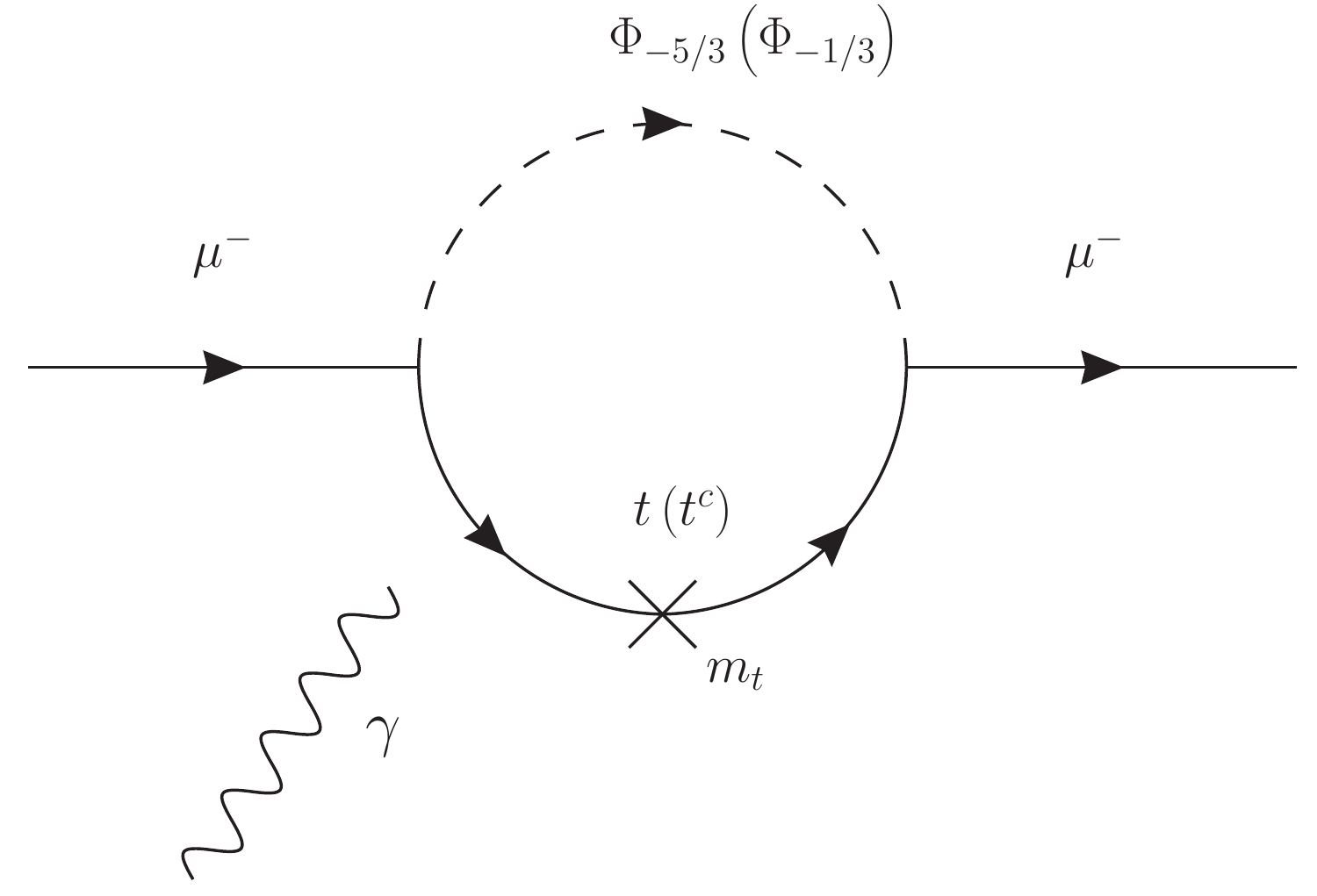}
\includegraphics[width=0.45\textwidth]{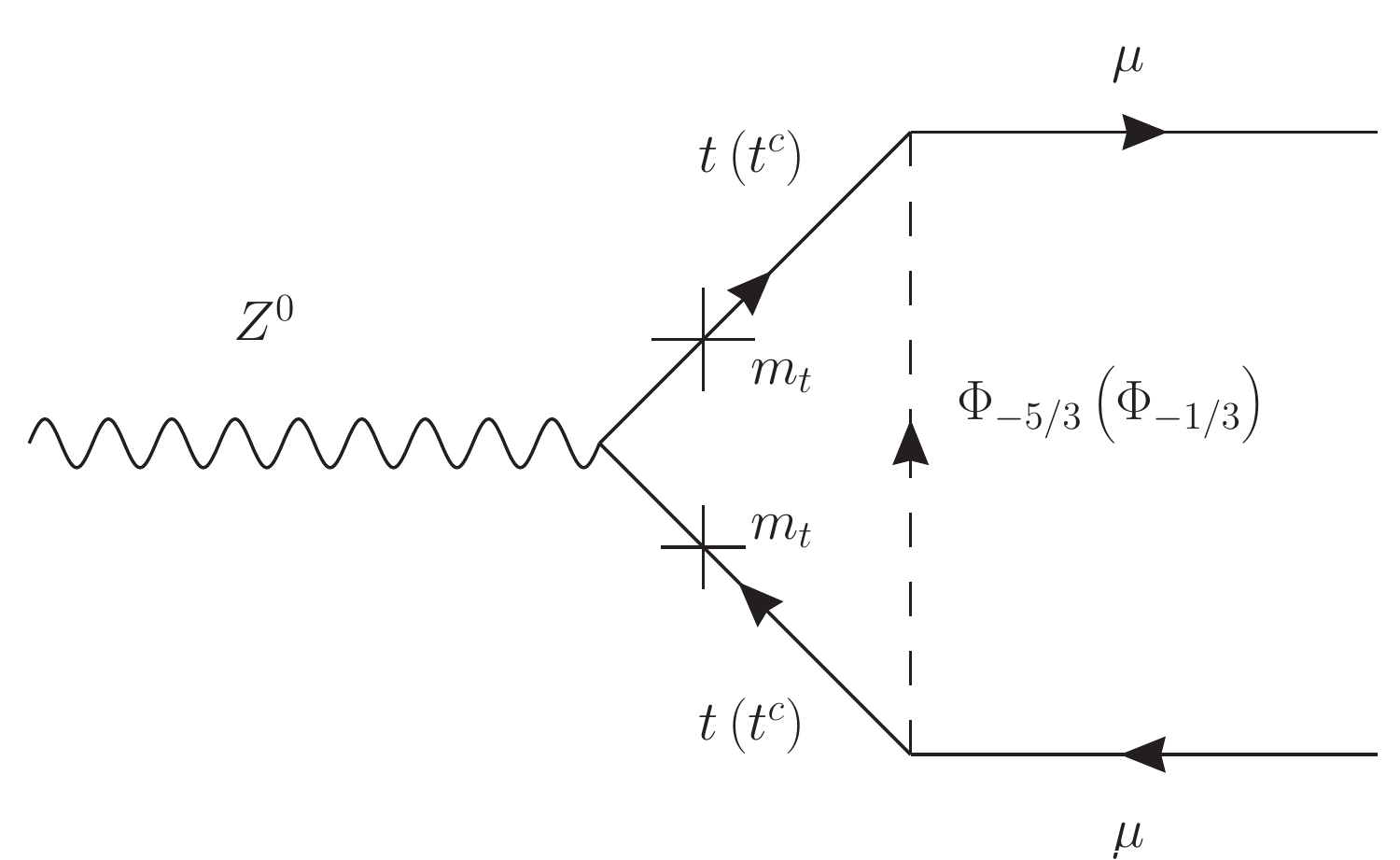}
\end{tabular}
\end{center}
\caption{Left: Feynman diagram showing the $m_t/m_\mu$ enhanced leptoquark contribution to the anomalous magnetic moment of the muon. Right: Leading correction to the $Z$ couplings to muons proportional to $m_t^2$.}         
\label{AMMfig}
\end{figure*}

An alternative explanation of the AMM of the muon involves leptoquarks~\cite{Djouadi:1989md,Davidson:1993qk,Couture:1995he,Chakraverty:2001yg,Cheung:2001ip,Mahanta:2001yc,Bauer:2015knc,Das:2016vkr,Biggio:2016wyy}. Here, even though the leptoquark must be rather heavy due to LHC constraints~\cite{Aaboud:2016qeg,ATLAS:2013oea,CMS-PAS-EXO-16-043,CMS-PAS-EXO-16-007,Khachatryan:2016jqo,Khachatryan:2016ycy,Aad:2016kww}, one can still get sizable effects in the AMM since the amplitude can be enhanced by $m_t/m_\mu$ compared to the SM. In fact, among the 5 scalar leptoquark representations which are invariant under the SM gauge group~\cite{Buchmuller:1986zs}, only two can in principle generate these enhanced effects as they possess couplings to left- and right-handed muons simultaneously:
\begin{itemize}
	\item $\Phi_1$: $SU(2)_L$ singlet with hypercharge $-2/3$.
	\item $\Phi_2$: $SU(2)_L$ doublet with hypercharge $-7/3$.
\end{itemize}
The corresponding Feynman diagram (see left diagram in Fig.~\ref{AMMfig}) involves a top quark providing the necessary chirality flip. 

A very similar diagram also contributes to $Z\to \mu^+\mu^-$. Since also this decay is sensitive to additional sources of chirality violation, again only the top contribution (shown in the right diagram in Fig.~\ref{AMMfig}) can be important. This relation among the two processes can be easily recognized in the effective field theory approach: at lowest dimension (dim-6) gauge invariance enforces that the effective interaction generating $a_\mu$ involves one coupling to the Higgs doublet while a modified $Z\to \mu^+\mu^-$ coupling requires couplings to two Higgs doublets. Interestingly, due to $SU(2)_L$ invariance, an explanation of the AMM of the muon with top couplings causes correlated effects in the decays of bottom mesons, i.e. $b\to s\nu\bar\nu$ ($b\to s\mu^+\mu^-$) for $\Phi_1$ ($\Phi_2$).

Furthermore, once leptoquarks couple to electrons and taus, additional NP effects in $a_e$, $a_\tau$, $Z\to \ell\ell^\prime$ and $\ell\to\ell^\prime\gamma$ are generated. Again, these effects will be correlated and can be used to identify the flavour structure of the leptoquark model under interest.

For all the processes discussed in this article, significant experimental progress can be expected in the future as there are many forthcoming and planned experiments: E989 will improve on the AMM of the muon~\cite{Grange:2015fou}, BELLE~II will open a new chapter in precision tau physics~\cite{Abe:2010gxa} improving the limits on $\tau\to\mu(e)\gamma$, $a_\tau$, but also on $B\to K^{(*)}\nu\nu$. Furthermore, the GigaZ experiment~\cite{Erler:2000jg} at a future ILC, or TLEP~\cite{Gomez-Ceballos:2013zzn} at the FCC, would produce order of magnitude more $Z$ gauge bosons than LEP. This will allow for extremely accurate determinations of $Z$ couplings to fermions and provide stringent limits on lepton flavour violating $Z$ decays. Last but not least, MEG II \cite{Baldini:2013ke} at PSI will significantly increase the existing $\mu\to e\gamma$ bounds and Mu3e the $\mu\to3e$ bounds~\cite{Blondel:2013ia}.

The goal in this article is to investigate the correlations among the AMM of the muon, $Z$ couplings to leptons, $\ell\to\ell^\prime\gamma$, $b\to s\nu\bar\nu$ and $b\to s\mu^+\mu^-$ and to discuss the implications for future experiments. In the next section we will present our model and provide the necessary formulas for the phenomenological analysis done in Sec.~\ref{analysis} before we conclude in Sec.~\ref{conclusions}.

\section{Model and observables}

As outlined in the introduction, a main motivation for this article is explaining the AMM of the muon via chirally enhanced loop effects. For leptoquarks the most pronounced enhancement stems from top quark loops. The two scalar leptoquarks ($\Phi _1$ and $\Phi _2$) which can generate these enhanced effects transform as $\Phi_1\!:\left( {3,1, - \frac{2}{3}} \right),\;{\Phi _2}\!:\left( {\bar 3,2, - \frac{7}{3}} \right)$ under the SM gauge group $SU(3)_c\times SU(2)_L\times U(1)_Y$. Their couplings to fermions are given by 
\begin{align}
L_{LQ}^{} =&\left( {\lambda _1^R\overline {{u^c}} \ell  + \lambda _1^L\overline {{Q^c}} i{\tau _2}L} \right)\Phi _1^\dag\nonumber\\
&+ \left( {\lambda _2^L\overline u L + \lambda _2^R\overline Q i{\tau _2}\ell } \right)\Phi _2^\dag  + {\rm{h}}{\rm{.c.}}\,.
\end{align}
Here we suppressed flavour indices. After electroweak symmetry breaking we expand the $SU(2)_L$ components and get 
\begin{align}
L_{LQ}^{} =&\left( {\overline {{u^c_f}} \left( {\lambda _{1,fi}^R{P_R} + \lambda _{1,fi}^L{P_L}} \right)\ell_i  - \tilde \lambda _{1,fi}^L\overline {{d^c_f}} {P_L}\nu_i } \right)\Phi _{- 1/3}^{ *}\nonumber\\
 &+ \left( {\lambda _{2,fi}^L\bar u_f{P_L}\nu_i  - \tilde \lambda _{2,fi}^R\overline d_f {P_R}\ell_i } \right)\Phi _{ - 2/3}^{*}\nonumber\\
 &+ \left( {\lambda _{2,fi}^L\bar u_f{P_L}\ell_i  + \lambda _{2,fi}^R\bar u_f{P_R}\ell_i } \right)\Phi _{- 5/3}^{ *}+ {\rm{h}}{\rm{.c}}.\,,
\end{align}
where the couplings $\tilde\lambda$ are related to $\lambda$ via CKM rotations and the $\Phi$ subscripts correspond to the electric charge. Since we are interested in couplings to top quarks only, we will from now on assume that all other couplings are zero and abbreviate the remaining ones via ${\lambda_{a,31}^{L,R}}\rightarrow{\lambda_e^{L,R}}$, ${\lambda_{a,32}^{L,R}}\rightarrow{\lambda_\mu^{L,R}}$ and ${\lambda_{a,33}^{L,R}}\rightarrow{\lambda_\tau^{L,R}}$ for $a=1,2$. Note that we dropped here the subscript $1$ and $2$ since we will consider the singlet case and the doublet case separately.

\subsection{Anomalous magnetic moments and radiative lepton decays}

The shifts in the AMMs are given by 
\begin{equation}
\delta {a_{{\ell _i}}} = \frac{m_{{\ell _i}}}{{4{\pi ^2}}}{\mathop{\rm Re}\nolimits} \left[ {C_R^{ii}} \right]\,,
\end{equation}
and the expressions for radiative lepton decays read
\begin{equation}
{\rm{Br}}\left( {{\ell _i} \to {\ell _f}\gamma } \right) = \frac{{{\alpha}{m_{{\ell _i}}^3}}}{{256{\pi^4}}}{\tau _{{\ell _i}}}\left( {{{\left| {C_L^{fi}} \right|}^2} + {{\left| {C_R^{fi}} \right|}^2}} \right)\,,
\end{equation}
with
\begin{align}
C_{L,\left({Q =  -1/3}\right) }^{fi}&=  - \frac{{{N_c}}}{{12}}\frac{{m_t}}{{{M^2}}}{\lambda _f^R\lambda _i^{L*}\left( {7 + 4\log \left( {\frac{{m_t^2}}{{{M^2}}}} \right)} \right)}\nonumber\\
&+\frac{N_c}{24M^2}\left(m_{\ell_i}\lambda_i^{R}\lambda_f^{R*}+m_{\ell_f}\lambda_f^{L}\lambda_i^{L*}\right)\,,\\
C_{L,\left( {Q =  - 5/3} \right) }^{fi}&=  \frac{{{N_c}}}{{12}}\frac{{m_t}}{{{M^2}}}{ { \lambda_{f}^R\lambda _{i}^{L*}\left( {1 + 4\log \left( {\frac{{m_t^2}}{{{M^2}}}}\right)} \right)}}\nonumber\\
&-\frac{3N_c}{24M^2}\left(m_{\ell_i}\lambda_f^{R}\lambda_i^{R*}+m_{\ell_f}\lambda^{L}_f\lambda^{L*}_i\right)
\,.
\end{align}
Here $M$ denotes the mass of the leptoquark which for $\Phi_2$ only differs among the $SU(2)$ components by $O(v^2/M^2)$. Furthermore, we set $N_c=3$. 

For the AMMs these expressions have to be compared to the experimental constraints for the electron~\cite{Giudice:2012ms} ($8\times 10^{-13}$), the muon (see \eq{muonAMMexp}) and the tau~\cite{Eidelman:2016aih} ($O(10^{-2})$). Concerning radiative lepton decays, the current $\mu \to e \gamma$ experimental bound, obtained by the MEG collaboration~\cite{TheMEG:2016wtm} is ${\rm Br}(\mu \to e \gamma) \leq 4.2 \times 10^{-13}$
and a future improvement of one order of magnitude is expected from the MEG II experiment~\cite{Baldini:2013ke}.
The BaBar collaboration~\cite{Aubert:2009ag} found only significant weaker bounds on radiative $\tau$ decays compared to the ones of the muon: the current experimental bound on $\tau \to \mu \gamma$ is $ {\rm Br}(\tau \to \mu \gamma) < 4.4 \times 10^{-8}$, while for $\tau \to e \gamma$ the limit is $ {\rm Br}(\tau \to e \gamma) < 3.3 \times 10^{-8}$, both at 90\% C.L.. Future sensitivities of the order of $10^{-9}$ are expected for these observables at BELLE~II~\cite{Bona:2007qt}.

\subsection{$Z$ couplings and decays}

The modifications of $Z$ couplings to leptons are given by
\begin{align}
\Delta \Gamma _{fi}^R &=  \pm \frac{{{g_2}N_c\lambda _f^{R*}\lambda_i^Rm_t^2\left( {1 + \log \left( {\frac{{m_t^2}}{{{M^2}}}} \right)} \right)}}{{32{\pi ^2}{c_w}{M^2}}}\,,\\
\Delta \Gamma _{fi}^L &=  \mp \frac{{{g_2}N_c\lambda _f^{L*}\lambda_i ^Lm_t^2\left( {1 + \log \left( {\frac{{m_t^2}}{{{M^2}}}} \right)} \right)}}{{32{\pi ^2}{c_w}{M^2}}}\,,
\end{align}
at leading order in $m_t^2/M^2$ and $m_Z^2/m_t^2$. Here the upper sign corresponds to $\Phi_1$ and the lower one to $\Phi_2$. The opposite signs reflect the fact that for $\Phi_2$ we have a top quark in the loop instead of an anti-top quark for $\Phi_1$. We constrain the flavour diagonal couplings from the combined fit presented in the PDG~\cite{Agashe:2014kda}. The constraints on the axial vector current are much stronger than the ones on the vector currents and are given by $-0.4\times 10^{-3}\leq\Delta\Gamma _{22}^R-\Delta \Gamma _{22}^L\leq0.7\times 10^{-3}$ for muons at the $1\,\sigma$ level. Note that here the interference with the SM for $\Phi_1$ is necessarily constructive while for $\Phi_2$ it is destructive. 

For $f\neq i$ the $Z\to\ell_i\ell_f$ branching ratios are given by
\begin{equation}
{\rm Br}\left( {{Z^0} \to \ell_i^ - \ell_f^+ } \right) = \frac{{g_2^2{m_Z}}}{{24\pi c_W^2}}\frac{1}{{\Gamma _Z^{\rm tot}}}\left( {{{\left| {\Delta \Gamma _{fi}^L} \right|}^2} + {{\left| {\Delta \Gamma _{fi}^R} \right|}^2}} \right)\,,
\end{equation}
with $\Gamma _Z^{\rm tot}=2.49\,$GeV~\cite{Agashe:2014kda}. The current bounds on $Z \to \ell_i^\pm \bar\ell_f^\mp$ by LEP are given by~\cite{Akers:1995gz},
\bea
{\rm Br}(Z \to \mu^\pm e^\mp)&<&1.7\times 10^{-6}\,,\\
{\rm Br}(Z \to \tau^\pm e^\mp)&<&9.8\times 10^{-6}\,,\\
{\rm Br}(Z \to \tau^\pm \mu^\mp)&<& 1.2\times 10^{-5}\,.
\eea
The future sensitivity of GigaZ to these LFV decays of the $Z$ boson
could reach an improvement of up to three orders of magnitude~\cite{AguilarSaavedra:2001rg}.

\begin{figure*}[tb]
\begin{center}
\begin{tabular}{cp{7mm}c}
\includegraphics[width=0.58\textwidth]{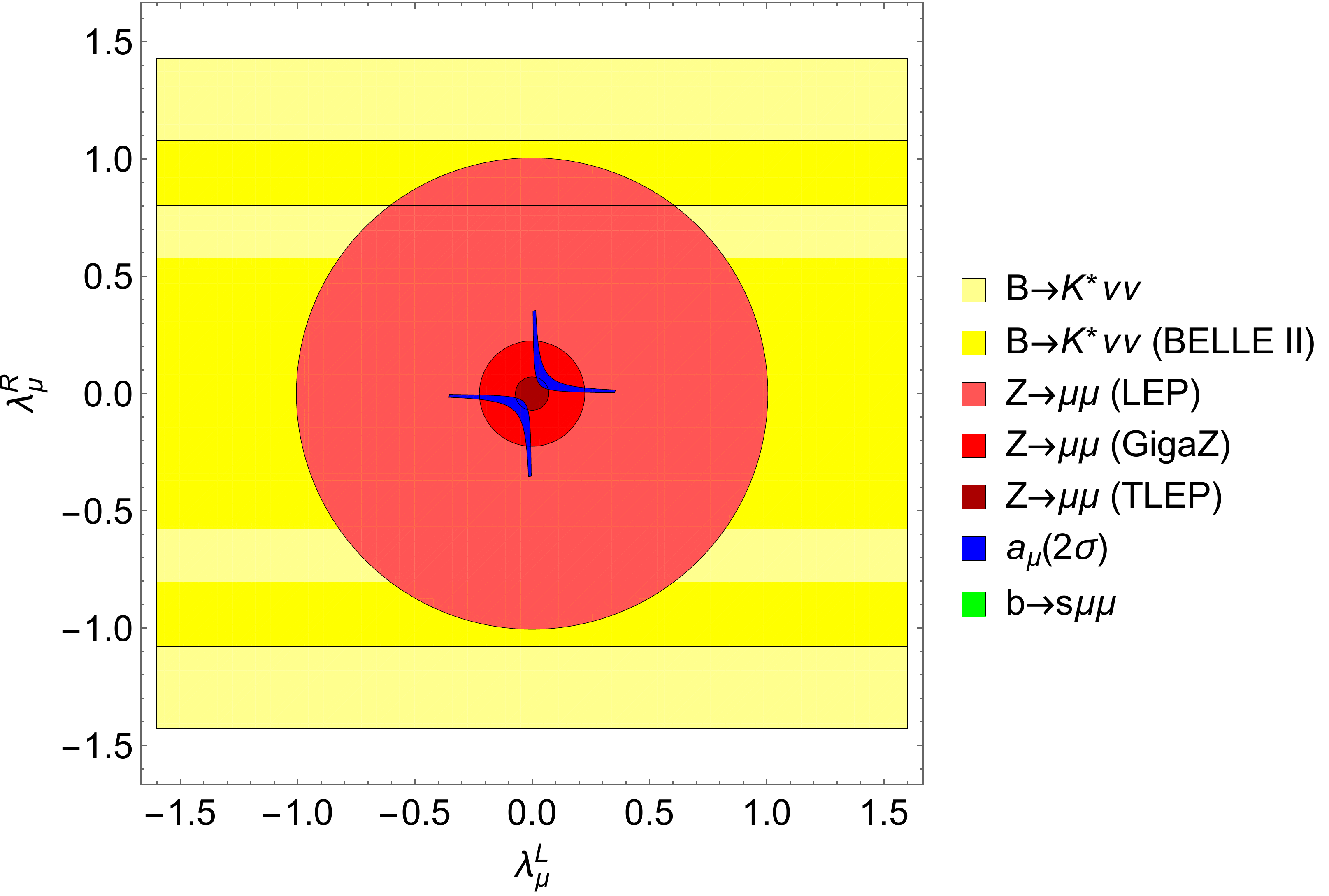}
\includegraphics[width=0.42\textwidth]{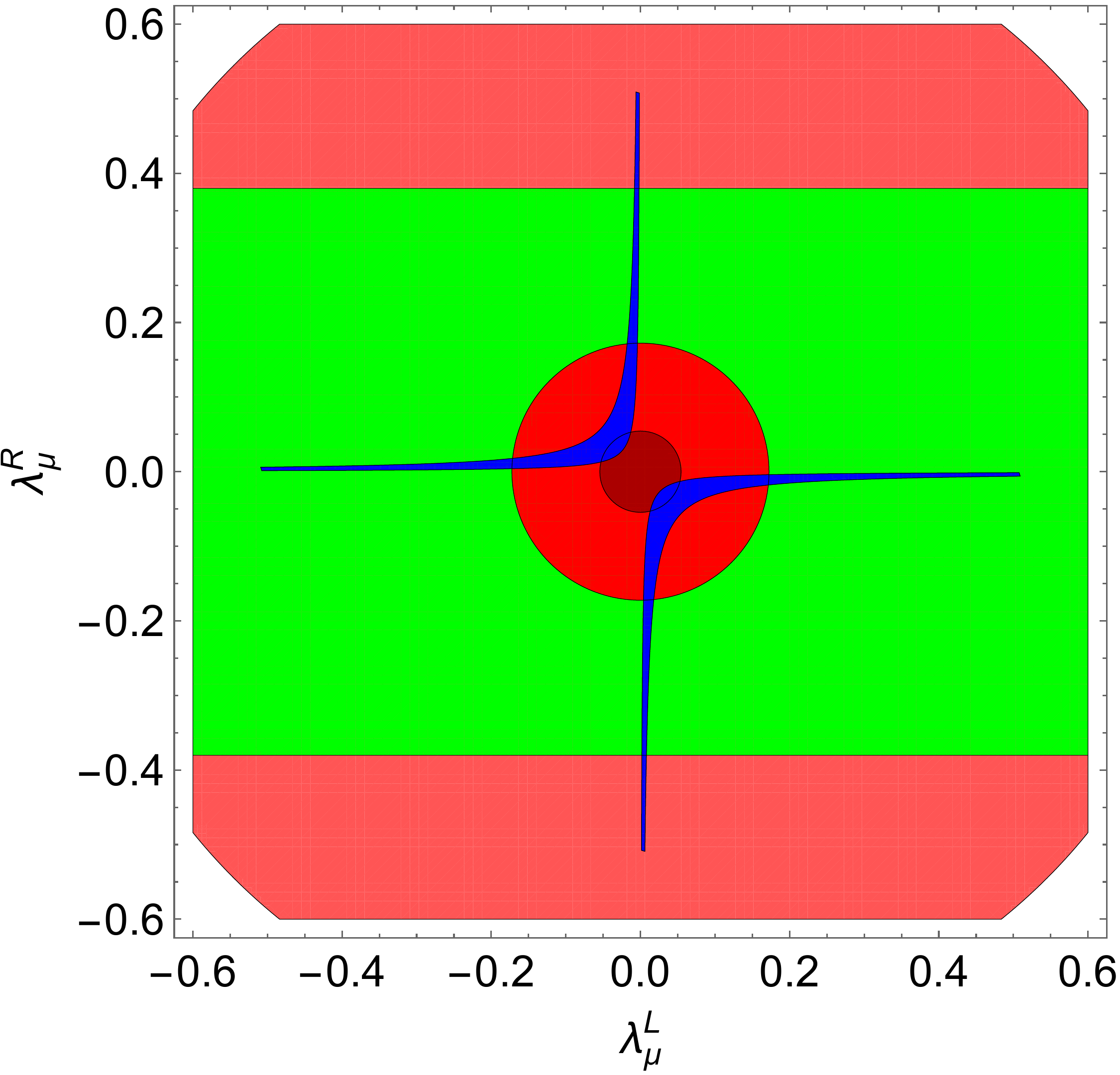}
\end{tabular}
\end{center}
\caption{Left: Allowed regions in the $\lambda^L_\mu$-$\lambda^R_\mu$ plane from current and future experiments for $SU(2)$ singlet leptoquarks $\Phi_1$ with $M=1\,$TeV. Right: Same as the left plot for the $SU(2)$ doublet leptoquark $\Phi_2$.}         
\label{muLmuR}
\end{figure*}

\subsection{$b\to s\ell^+\ell^-$ transitions}

Only the $SU(2)$ doublet leptoquark $\Phi_2$ contributes to $b\to s\ell^+\ell^-$. Using the effective Hamiltonian
\begin{align}
H_{\rm eff}^{\ell_f\ell_i}=- \dfrac{ 4 G_F }{\sqrt 2}V_{tb}V_{ts}^{*} \sum\limits_{a = 9,10} C_a^{fi} O_a^{fi}\,
,\nn {O_{9(10)}^{fi}} =\dfrac{\alpha }{4\pi}[\bar s{\gamma ^\mu } P_L b]\,[\bar\ell_f{\gamma _\mu }(\gamma^5)\ell_i] \,,
\label{eq:effHam}
\end{align}
we have
\begin{equation}
C_9^{fi} = C_{10}^{fi} = \frac{{ - \sqrt 2 {\pi}}}{{4{G_F}{\alpha}{M^2}}}\lambda_f^{R*}\lambda_i^{R}\,.
\end{equation}
Even though one cannot fully explain the observed anomalies in $b\to s\mu^+\mu^-$ transitions with $C_9^{fi} = C_{10}^{fi}$, a slight improvement of $1\,\sigma$ in the global fit compared to the SM is possible and the preferred $2\,\sigma$ range is given by $-0.64 \leq C_9^{22} = C_{10}^{22}\leq 0.33$~\cite{Altmannshofer:2014rta,Altmannshofer:2015sma} (see also Ref.~\cite{Descotes-Genon:2015uva,Hurth:2016fbr}).

\subsection{$B\to K^{(*)}\nu\bar{\nu}$}

Only the $SU(2)$ singlet leptoquark $\Phi_1$ contributes to $b\to s\nu\bar\nu$ transitions at tree-level. Following Ref.~\cite{Buras:2014fpa} we write the relevant effective Hamiltonian as
\begin{align}
H_{\rm eff}^{\nu_f\nu_i} = -\frac{{4{G_F}}}{{\sqrt 2 }}{V_{tb}}V_{ts}^*{C_L^{fi}}{O_L^{fi}}\,,\\
O_{L}^{fi}= \frac{\alpha }{{4\pi }} [\bar s{\gamma ^\mu }{P_{L}}b][{{\bar \nu }_f}{\gamma _\mu }\left( {1 - {\gamma ^5}} \right){\nu _i}]\,,
\end{align}
with
\begin{equation}
C_L^{fi} = \frac{{\sqrt 2 {\pi}}}{{4{G_F}{\alpha}{M^2}}}\lambda _f^{L*}\lambda _i^L\,,
\end{equation}
and $C_L^{\rm SM}\approx-1.47/s_w^2$. The branching ratios normalized by the SM predictions read
\begin{equation}
{R_{K^{(*)}}^{\nu\bar{\nu}}} = 
\frac{1}{3}\sum\limits_{f,i=1}^3 \dfrac{ \left|  {C_L^{fi}} \right|^2}{\left| {C_L^{\rm SM}} \right|^2} \,.
\end{equation}
This has to be compared to the current experimental limits ${R_K^{\nu\bar{\nu}}} < 4.3$~\cite{Lees:2013kla} and ${R_{{K^*}}^{\nu\bar{\nu}}} < 4.4$~\cite{Lutz:2013ftz}. The Belle II sensitivity for $B\to K^{(*)}\nu\bar{\nu}$ is 30\% of the SM branching ratio~\cite{Abe:2010gxa}.

\subsection{$h\to \ell\ell^\prime$}

Let us finally consider the effects in lepton flavour violating Higgs decays. Here we have~\cite{Dorsner:2016wpm} 
\begin{align}
\label{eq:LQ1-BR}
 &{\rm Br}(h \to \ell_f \ell_i)=\\
&\qquad\frac{1}{\Gamma_h}\left(\frac{9 m_{h}m_{t}^{2}}{2^{13}\pi^{5}v^{2}}\left(\left|\lambda_{f}^{L}\lambda_{i}^{R}\right|^{2}+\left|\lambda_{f}^{R}\lambda_{i}^{L}\right|^{2}\right)\left|g_1(M_{\phi})\right|^{2}\right)\,.\nonumber
\end{align}
However, this equation should only be considered as an order of magnitude estimate, since also quartic couplings of two leptoquarks to two Higgses are allowed by renormalizability which would contribute to this process. Furthermore, the possible rates are far below future sensitivities and we will not discuss this decay in our phenomenological analysis.

\section{Phenomenological analysis and future prospects}
\label{analysis}

In a first step, let us consider the case in which the leptoquark couples to muons only. In this setup constraints from the AMM of the muon, $Z\to\mu^+\mu^-$ and $b\to s\mu^+\mu^-$ (for $\Phi_2$) or $b\to s\nu\bar\nu$ (for $\Phi_1$) arise. The allowed regions from these processes are shown in Fig.~\ref{muLmuR} for a leptoquark mass of 1~TeV. Note that, neglecting logarithmic effects, the constraints on the couplings simply scale like $M/(1\,\rm{TeV})$. For $Z\to\mu^+\mu^-$ also the expected future bounds at GigaZ are shown where one can expect an increase of precision by a factor of around 20~\cite{Baer:2013cma}. Also the projected TLEP bounds and BELLE~II limits for $B\to K^{(*)}\nu\bar{\nu}$ are depicted.

\begin{figure*}[tb]
\begin{center}
\begin{tabular}{cp{7mm}c}
\includegraphics[width=0.55\textwidth]{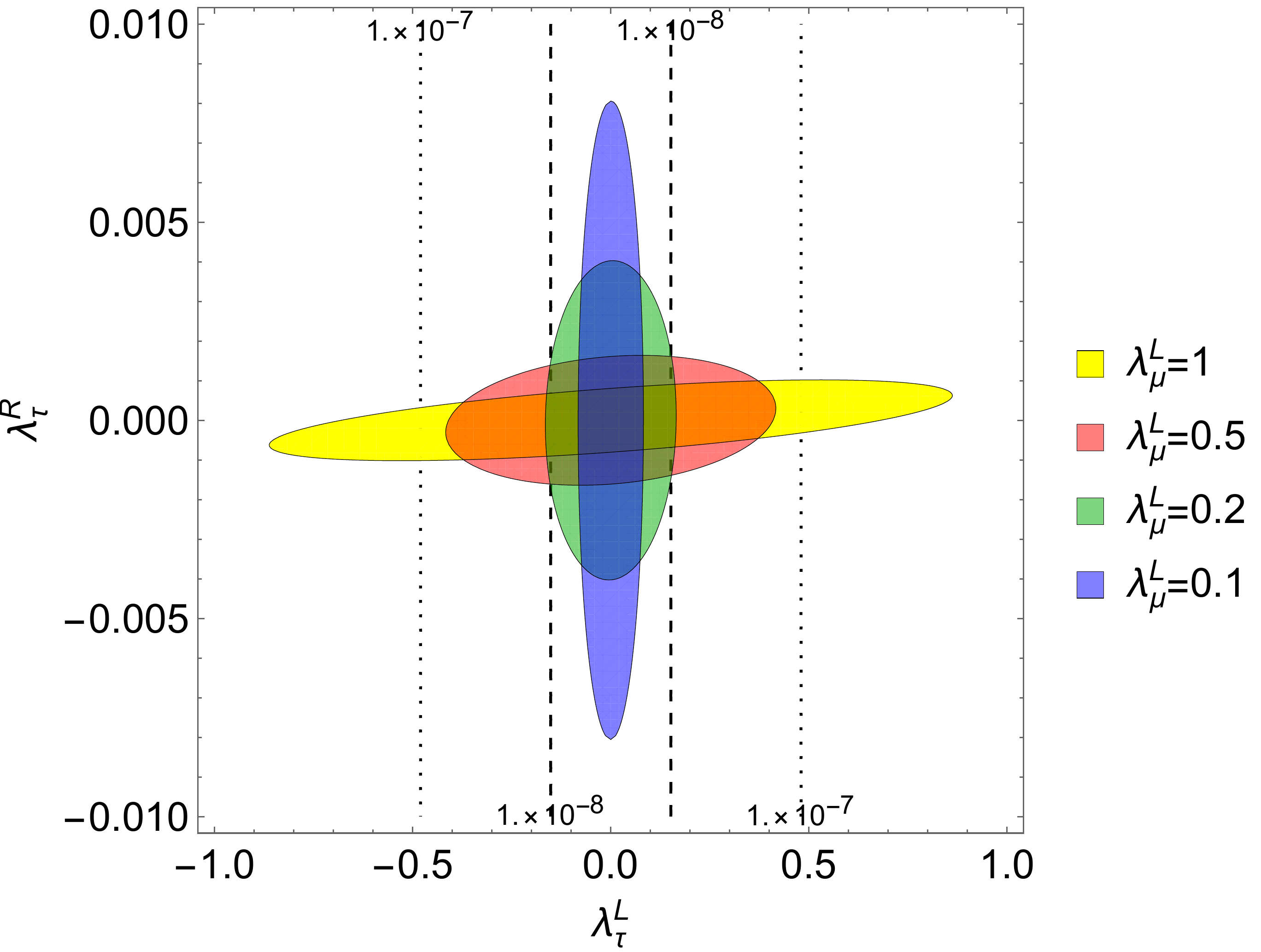}
\includegraphics[width=0.45\textwidth]{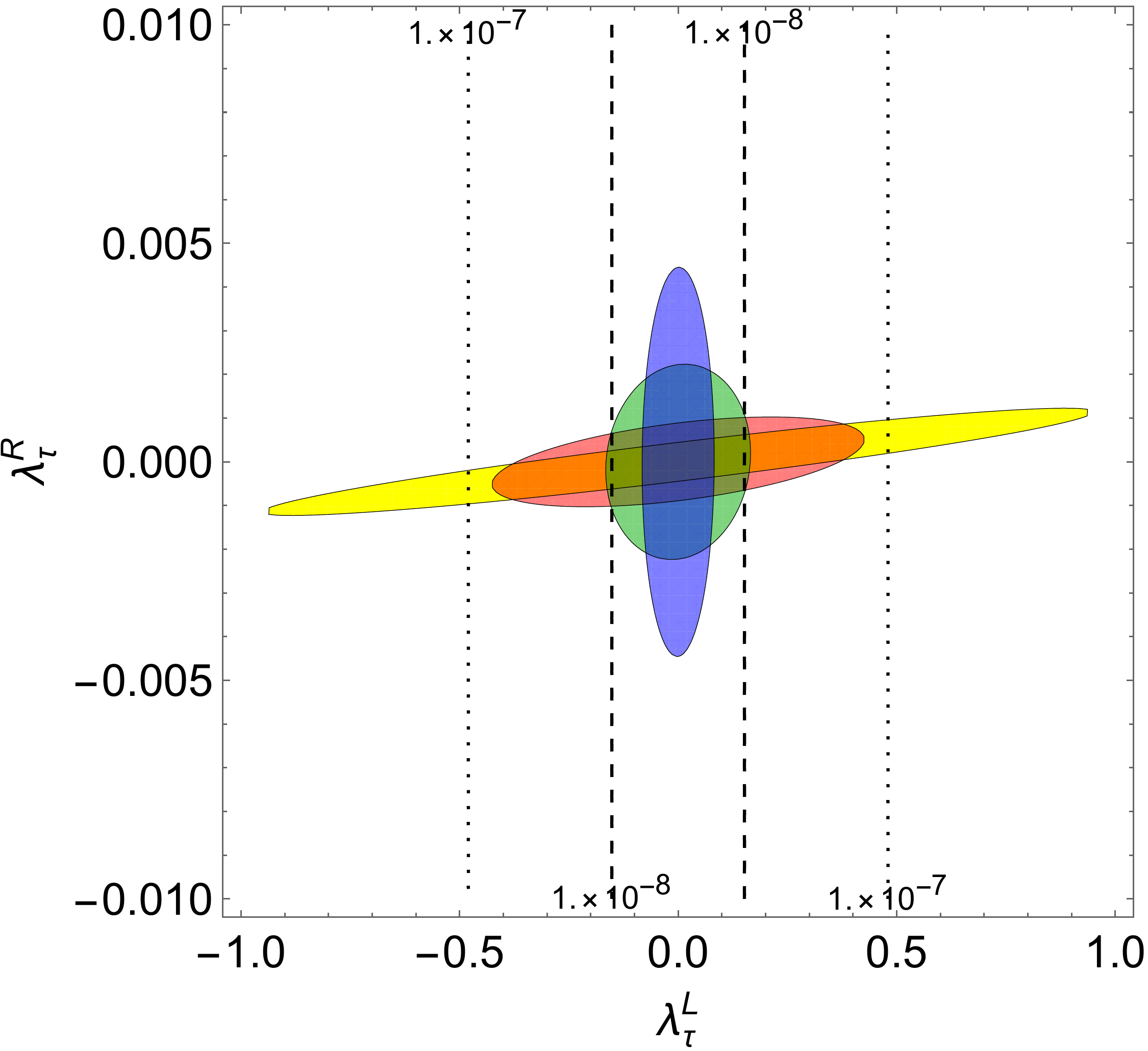}
\end{tabular}
\end{center}
\caption{Left: Allowed regions from $\tau\to\mu\gamma$ in the $\lambda^L_\tau$-$\lambda^R_\tau$ plane for different couplings to left-handed muons assuming vanishing couplings to electrons for $\Phi_1$ and $M=1\,$TeV. The right-handed coupling to muon is fixed by requiring $\delta a_\mu=10^{-9}$. The contour lines denote the predicted branching ratio for $Z\to\tau^\pm\mu^\mp$. Right: Same as left plot for for the $SU(2)$ doublet leptoquark $\Phi_2$.}         
\label{tauLtauR}
\end{figure*}

Once couplings to tau leptons are present, the situation gets more involved. Assuming that one aims at addressing the AMM of the muon, non-zero branching ratios for $\tau\to\mu\gamma$ and $Z\to\tau^\pm\mu^\mp$ are generated once tau couplings are turned on. This is illustrated in Fig.~\ref{tauLtauR} where the complementary constraints on the couplings for both leptoquark representations are shown. Note that couplings to tau leptons can only be sizable if there is simultaneously a large hierarchy $\lambda^R\ll\lambda^L $ or $\lambda^L\ll\lambda^R$ both for muon and tau couplings.

Since the bounds from $\mu\to e\gamma$ are so stringent, one can see without detailed analysis that couplings to electrons must be extremely tiny if couplings to muons are sizable. In fact, $\mu\to e\gamma$ rules out any observable effect in the AMM of the electron if there is a deviation in the anomalous magnetic moment of the muon as currently suggested by theory and experiment. Furthermore, no measurable effect in $Z\to \mu e$, even at future colliders, is possible since the bounds on the $Z\mu e$ coupling from $\mu\to 3e$ are much more stringent~\cite{Crivellin:2013hpa}.

In case of couplings to electrons and taus but vanishing couplings to muons (i.e. giving up an explanation for $(g-2)_\mu$), the situation is similar to the one with couplings to muon and taus. However, the constraints from the AMM of the electron are stronger, allowing only for smaller couplings to electrons and therefore leaving more freedom in tau couplings.

\section{Conclusions}\label{conclusions}

In this article we considered the impact of leptoquarks on the AMM of the muon, $Z$ couplings to leptons, radiative lepton decays, $b\to s\ell^+\ell^-$ and $b\to s\nu\bar\nu$. There are two leptoquark representations which are phenomenologically interesting since they can give $m_t/m_\mu$ enhanced effects in the AMM of the muon and potentially explain the observed anomaly. Their main features, which allow to distinguish them also using low-energy precision experiments, are:

\begin{itemize}
	\item $SU(2)$ singlet $\Phi_1$: Constructive interference with the SM in axial vector current contributing to $Z\to\ell\ell$ and effects in $b\to s\nu\bar{\nu}$.
	\item $SU(2)$ doublet $\Phi_2$: Destructive interference with the SM in axial vector current contributing to $Z\to\ell\ell$ and effects in $b\to s\mu^+\mu^-$.
\end{itemize}

In our numerical analysis we found that if the AMM of the muon is explained by one of these leptoquarks, future experiments like TLEP or GigaZ should be able to see deviations in $Z\to\mu^+\mu^-$. If one aims at an explanation of the AMM of the muon, stringent constraints on couplings to electrons and taus arise. Due to the constraints from $\mu\to e\gamma$, couplings to electrons must be zero to a good approximation but there is still space for sizable couplings to taus, allowing for measurable effects in $Z\to\tau^+\tau^-$. However, also here sizable couplings are only possible in the presence of a hierarchy between the left- and right-handed couplings both in the muon and in the tau sector. In this case, sizable rates for $\tau\to \mu\gamma$ are generated and also $Z\to \tau^\pm\mu^\mp$ could be in the range of future precision experiments.


{\it Acknowledgments} --- {\small A.C. and D.M. are grateful to Michael Spira for very useful discussions. The work of A.C. and D.M. is supported by an Ambizione Grant of the Swiss National Science Foundation. G.D. thanks Gino Isidori, Riccardo Barbieri and Paride Paradisi for helpful discussions.}

\bibliography{BIB}

\end{document}